# Physics-informed genetic algorithms (PIGAs) facilitating LIBS spectral normalization with shockwave characteristics

Ying Zhou,[1] Jian Wu,[1, a)] Mingxin Shi,[1] Minxin Chen,[1] Jinghui Li,[1] Xinyu Guo,[1] Yuhua Hang,[2] Cuixiang Pei,[3] and Xingwen Li [1]

[1)] *State Key Laboratory of Electrical Insulation and Power Equipment, Xi'an Jiaotong University, Xi'an 710049, China*
[2)] *Suzhou Nuclear Power Research Institute, Suzhou 215004, China*
[3)] *State Key Laboratory for Strength and Vibration of Mechanical Structures, Shaanxi Engineering Research Center of NDT and Structural Integrity Evaluation, Xi'an Jiaotong University, Xi'an, 710049, China*

[a)] Author to whom correspondence should be addressed: jxjawj@mail.xjtu.edu.cn

Inspired by physics-informed neural networks (PINNs) inheriting both the interpretability of physical laws and the efficient integration capability of machine learning, we propose a framework based on stoichiometric ablation for LIBS spectral normalization, encoding physical constraints between LIBS intensities and shockwave characteristics (temperature $T_{\text{shock}}$ and pressure $P$) into optimization algorithms with multiple independent objectives, named physics-informed genetic algorithms (PIGAs). It is characterized by its applicability to the wider laser energy range covering laser-induced breakdown to significant plasma shielding and spectral lines undergoing self-absorption outperforming the widely-used physical linear or multivariate data-driven normalization methods. The home-made end-to-end LAP-RTE codes serves as the benchmark to validate the physical reciprocal-logarithmic transformation and its extensibility to self-absorption spectral lines for PIGAs. Next experimental spectral lines are statistically used to validate PIGAs' correction effects, the median RSDs of spectral intensities can be effectively reduced by 85% (corrected by $P$) and 88% (corrected by $T_{\text{shock}}$) for 108 Fe I lines, while for 33 Fe II lines, reduced by 77% (corrected by $P$) and 86% (corrected by $T_{\text{shock}}$). Seventeen self-absorption lines are also corrected effectively, with RSDs being reduced by 78% (corrected by $P$) and 89% (corrected by $T_{\text{shock}}$). Our proposed idea of combining optimization methods to quantify unknown parameters in normalization strategies can also be extended to excavate the correlation between parameters for other low-temperature plasma fields with similar processes.

Laser-induced breakdown spectroscopy (LIBS), known as the "future superstar" of analytical atomic spectroscopy[1], can provide information on material composition based on laser-target interaction[2], with its advantages of fast, no sample preparation and multi-elemental analyses[3,4]. However, it has still been "suffering" criticisms including the process being irreproducible[5] and signals yielded unstable[6,7], because of its strongly nonlinear, fleeting nature[8,9] and high sensitivity to any random or regular changes in laser-target parameters[10], such as laser irradiances[11].

There has been a growing awareness to monitor other experimentally feasible plasma signals as external references to normalize LIBS signals[12], which can follow laser-target interaction but appear to be composition independent, such as crater volume[13], plasma image[14,15] and acoustic signals[16]. Wherein the acoustic signal has been receiving more attention for its ease of mounting with microphones, correlating with information about laser-material interaction including the amount of ablated materials and laser irradiances[17,18].

Two typical types have been proposed for acoustic signal normalization. The first univariate analysis uses the acoustic amplitude, energy or magnitude in frequency domain for linear normalization[19], which has been found to be proportional to the ablation mass and laser irradiance in low laser irradiation regions. However, such linear normalization was reported to be insignificant in wider laser irradiation ranges when laser-induced breakdown or plasma shielding occurs, also for spectral lines undergoing self-absorption[20,21]. The second multivariate scheme, based on machine learning, establishes regression between LIBS intensity and acoustic intensity matrixes over a period of time[22], but the lack of physical constraints makes it plagued with under/over-fitting problems[23,24]. Besides that, the acoustic signal, as part of the initial strong shockwave, has been attenuated showing more susceptible to environmental interferences.

To address these issues, we focus on directly measuring shockwaves and propose a normalization framework encoding the physical correlation between LIBS intensity and shockwave characteristics into machine learning for LIBS spectral normalization, which is named physical-informed genetic algorithms (PIGAs). In this framework, direct physical correlation guarantees the validity for situations involving different spectral fluctuation-inducing sources and saturated plateaus over wider ranges of laser irradiance, also for spectral lines experiencing self-absorption. While the genetic algorithm with multi-objectives are used to determine unknown parameters in the above-mentioned physical correlation.

This letter is organized as two sections covering theoretical derivation of the correlation, followed by cross-validating the effects of PIGAs with experiments and physical modeling describing the complete coupled progress of laser-material interaction. One example for laser irradiance fluctuating is used in validation to illustrate the framework's processing flow and its correction results.

In the first section, we derive the physical correlation between LIBS intensity and shockwave characteristics to introduce the framework of PIGAs. In local thermodynamic equilibrium (LTE), for an isothermal, homogeneous plasma, spectral intensity $I$ can be obtained from the radiation transport equation[25]

$$I(\nu)=B(\nu)\left(1-e^{-\tau(\nu)}\right)=B(\nu)\tau(\nu)(\text{when}\quad \tau(\nu)\rightarrow 0), \tag{1}$$

where $B$ is the Planck function and $\tau$ is the optical depth.

When $\tau$ tends to zero, $I$ can be written as the product of $B$ and $\tau$. Then Eq.(1) can be extended

$$I\,(\nu)=\left(\frac{\pi e^2 f}{m_e c}\, g_{\text{low}}\, \frac{2h\nu^3}{c^2}\right) N\left(\frac{x_i}{Z_i(T_e)}\, e^{-E_{\text{up}}/k_B T_e}\right)$$
$$=CN\left(\frac{x_i}{Z_i(T_e)}\, e^{-E_{\text{up}}/k_B T_e}\right)=CNI_{\text{tem}}, \tag{2}$$

where C is a constant for a specific line, $N$ is the total particle number density of a given element A in the plasma, $x_i$ is the proportion of A at $i$-th ionization state, $Z_i(T_e)$ is the partition function at the corresponding plasma temperature $T_e$, here we define the last factor in Eq.(2) as $I_{tem}$.

In stoichiometric equilibrium, $N$ is proportional to the concentration $c_A$ in the sample, suggesting that when $I_{tem}$ keeps constant, $I$ shows proportional to $c_A$, which is the basis for LIBS to measure elemental concentration. However, in actual experiments, potential bias in parameters such as laser irradiances, tends to make $I_{tem}$ continuously changing, which in turn makes the corresponding spectra fluctuating even when $c_A$ keeping constant, therefore we need to quantify $I_{tem}$ in detail. It is clear that the first two terms $x_i$ and $Z_i(T_e)$ in $I_{tem}$ obey the well-known Saha equation and the Boltzmann distribution in LTE[26], which are determined by $T_e$ when $N$ holds constant. The last term $\exp(-E_{up}/k_B T_e)$ also presents in an exponential form of $T_e$. These findings suggest that for a certain $c_A$, $I$ determined by $I_{tem}$, is only controlled by $T_e$. Therefore, we continue to analyze how shockwave characteristics correlate with $T_e$ and hence reflect spectral intensities.

We propose $T_{shock}$ as the temperature at the shockwave front. Since the shockwave front and plasma behind share the continuous medium, we can correlate $T_e$ and $T_{shock}$ through the heat transfer equation and motion constraints from Taylor-Sedov strong explosion. After that they can be converted

$$T_{shock}=T_e+(T_0-T_e)\times erf\left(\eta_{shock}(\tau)\right), \quad (3)$$

where $T_0$ represents the ambient temperature, $erf(\eta_{shock})$ is the error function of the similarity variable $\eta_{shock}$ determined by the shockwave propagation time $\tau$ (detailed derivation attached in Supplementary Material).

Therefore, we can measure $T_{shock}$ to determine $T_e$ and thus follow the variation of $I$. However, it is not feasible to measure $T_{shock}$ directly in experiments whereas the pressure $P$ can be gauged readily through microphones.

So, in the following, we try to correlate $T_{shock}$ and $P$. For the ideal gas, we apply the commonly-used Rankine-Hugoniot relation[27]

$$\frac{T_{shock}}{T_0}=\frac{P}{P_0}\frac{(\gamma_0+1)+(\gamma_0-1)(P/P_0)}{(\gamma_0-1)+(\gamma_0+1)(P/P_0)}, \quad (4)$$

where $P_0$ and $\gamma_0$ are the ambient pressure and the ratio of the specific heat.

Since the pressure at the shockwave front is generally greater than the ambient pressure, i.e., $P/P_0 >> 1$, then Eq.(4) can be simplified

$$T_{shock}=\frac{T_0}{P_0}\frac{(\gamma_0-1)}{(\gamma_0+1)}P, \quad (5)$$

which suggests $P$ to be linearly correlated with $T_{shock}$. Ref.28 and 29 also reported the linear relationship between $P$ and $T_{shock}$.

From the above derivation and discussion, we have elaborated that two dependent shockwave characteristics ($T_{shock}$ and $P$) can capture spectral intensities by directly correlating with $T_e$. Since $T_e$ stays in reciprocal-exponential transformation in Eq.(2), it might be more reasonable to relate the natural–logarithmic of $I$ to the reciprocal of $T_{shock}$(or $P$) instead of their original forms directly. Ref.30 also confirmed that ln$I$ yielded more efficient compared to normal $I$ in spectral preprocessing.

As for the correlation algorithms in PIGAs, we define the form of physical correlation between $I$ and $T_{shock}$(or $P$) as

$$\ln I=\sum_{n=0}^{q} a_n \left(\frac{1}{x}\right)^n, \ x=T_{shock} \text{ or } P, \quad (6)$$

where the degree $q$ and coefficients $a_n$ are searched through a controlled elitist genetic algorithm with multiple objectives[31]. It uses a population-based approach to generate more than one solution in each iteration. Four main operators (selection, crossover, mutation, and elite-preservation) are applied to make the new population close to the Pareto-optimal front. This procedure differs from classical optimization algorithms where one point is generated in each iteration.

Multiple objectives yield tradeoffs with one another avoiding biases from prior weights for hyperparameters in single-objective optimization[32]. For specific objectives:

For the degree $q$, Firstly it should be sufficient to ensure that $T_{shock}$(or $P$) can adequately follow the change in $I$. The second is to prevent over-fitting, the range of $T_{shock}$(or $P$) should be extended outward by 10% and make sure that within this range the predicted intensity $I_{pre}$ shows no unreasonable such as sudden jumps or unphysical data.

For coefficients $a_n$, three objectives must be satisfied simultaneously: 1) $R^2$ between the actual fluctuating $I_{actual}$ and the predicted $I_{pre}$ should be maximized; 2) percentage relative error RE(%) between $I_{actual}$ and $I_{pre}$ should be minimized; and 3) RSD of $I_{actual}/I_{pre}$ should be minimized.

For data obtaining of $I_{actual}$ and $T_{shock}$(or $P$), for physical modeling, they are directly from calculations; and for experiments, $I_{actual}$ and $P$ are from direct measurements, while $T_{shock}$ goes through back-calculating from the measured $P$ based on the assumption in Eq.(5), i.e., $T_{shock}=m_1P+m_2$. Variables ($m_1$, $m_2$) share the same objectives for $a_n$.

Therefore, once $q$ and $a_n$ are determined in Eq.(6), we can get a predicted $I_{pre}$ through its correlation with shockwave characteristics $T_{shock}$(or $P$) and use it to normalize the actual $I_{actual}$, i.e., $I_{actual}/I_{pre}$. We name the above normalization framework as PIGAs with its features of physical constraints on reciprocal-logarithmic transformation and optimization algorithms to determine unknown parameters within it.

In the second section, we take the laser irradiance variation as one example to validate the spectral correction effects of PIGAs especially for saturated plateaus over the wider irradiance ranges and spectral lines undergoing self-absorption, when dealing with two types of data sources, i.e., numerical modeling and experiment. Here we choose pure iron instead of mixtures as the sample to keep $c_A$ constant and avoid elemental fractionation[33].

For the acquisition of calculation data, we develop home-made end-to-end laser-ablation plasma-radiation transport equation (LAP-RTE) codes to describe the whole process of laser ablation, plasma expansion, ionization and spectral radiation for pure iron under nanosecond pulsed laser interaction in an air-like atmosphere with 78% $N_2$ and 22% $O_2$(more details attached in Supplementary Material). Typical laser pulse is set in gaussian shape with 1064-nm wavelength, 10-ns pulse width. Peak irradiances are divided into ten equal groups between $3.875\times10^{12}$ W/m$^2$ and $5\times10^{12}$ W/m$^2$, which covers the range from laser-induced breakdown just emerging to significant plasma shielding happening. The line intensities are integrated from 400 ns up to 4.5 μs to simulate the typical temporal window of the spectrometer in experiments.

For the measurement of experimental data, a Nd:YAG laser beam (Lapa 80, BeamTech) operating at 1064-nm wavelength, 10-ns pulse width and 1-Hz repetition rate was focused by a convex lens (f = 30 mm) onto the fresh location of the sample (ultra-low carbon pure iron, YSBS11058-2021, NCS Testing Technology Co. Ltd.) at a normal angle of incidence (Figure 1(a)). Spectral emission was coaxially collected into a fiber optic spectrometer (AvaSpec-ULS2048-10-USB2-RM Multi Channel, 180-1060 nm spectral range, UF/UE grating, 2400-3600 lines/mm, 0.05–0.09 nm resolution, 1.05 ms minimum gate width) with 400 ns gate delay. The first 6 channels were used and calibrated with a standard light source (AvaLight-Hal, Avantes, Netherlands), whose cross regions were avoided. A pressure sensor (113B26, PCB Piezotronics, Inc) was mounted at 45° and 4.3 mm (consistent with the calculation selection) away from the ablation point to measure the shockwave. The incident laser energy measured by an energy meter (Nova II, P/N 7Z01550, Ophir), was uniformly set in 11 groups between 18 mJ and 28 mJ to generate irradiances close to the calculation, yielding a 750 μm beam diameter measured by a beam-profiling camera (SP620U, Spiricon) and an 830 μm ablation crater diameter by an optical microscope (bx53m, Olympus Instruments). The 5th shot was collected for each spectrum and the corresponding shockwave profile preventing deep-crater effects, and 20 repetitions were performed for each laser energy configuration.

To quantitatively follow the statistical properties behind, a total of 108 Fe I and 33 Fe II lines are captured to cross-validate both in calculations and experiments from 100 nm to 830 nm. McWhirter criterion are used to check the LTE state.

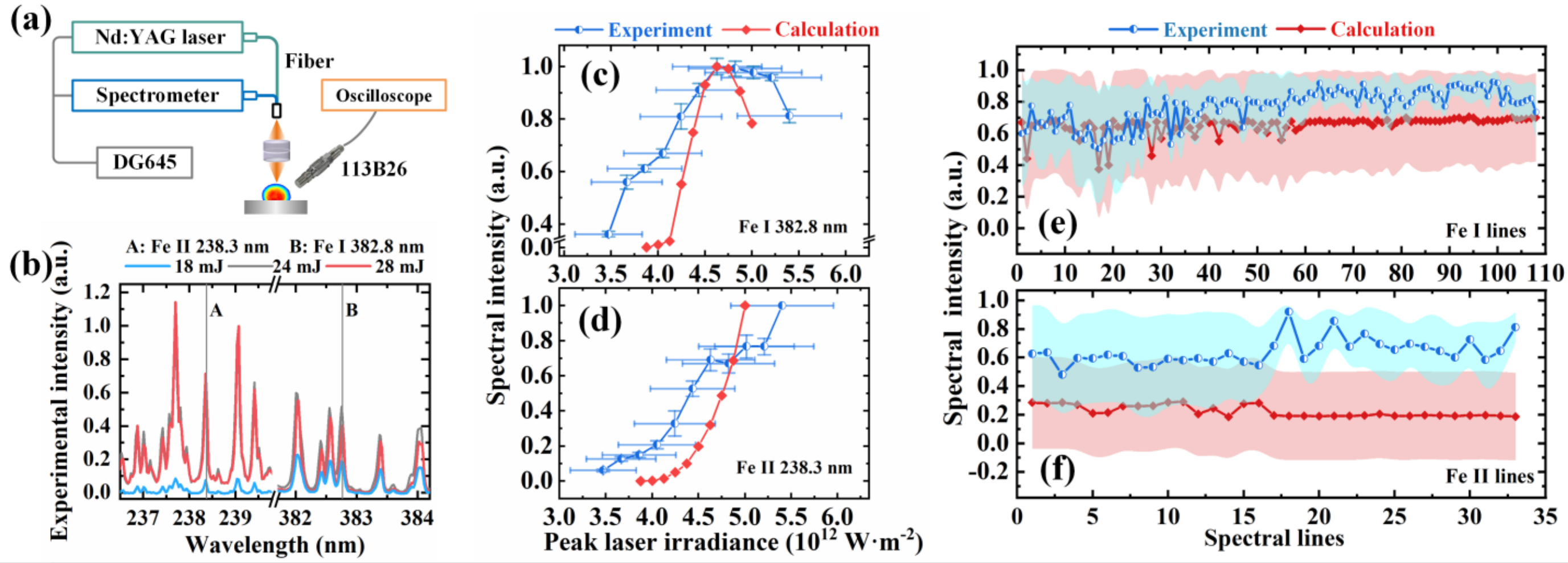

**FIG. 1** Spectral intensity changes induced by laser irradiance variation. (a)Experimental setup. Evolution of (b)experimental spectral intensities with laser energies, line intensities with laser irradiances for (c)Fe I 382.8 nm and (d)Fe II 238.3 nm. Specific ranges for (e)108 Fe I lines and (f)33 Fe II lines.

Figure 1(b) illustrates the experimental spectral changes. We find the unimodal trend of 108 Fe I line peak intensities (also found in Ref.34 and 35) and monotonic trend of 33 Fe II line peak intensities (also reported in Si II lines[36]) with laser irradiance increasing, taking Fe I 382.8 nm (Figure 1(c)) and Fe II 238.3 nm (Figure 1(d)) as examples, where the intensity errors come from 20 repetitions and irradiance errors indicate the inconsistency between the spot and ablation crater sizes. The specific change for each line is further shown in Figure 1(e)-(f). It is clear that these strong variations corresponding to only one concentration collapse the linear relationship between the spectral intensity and the element concentration.

Next, calculated spectral lines are corrected by PIGAs using the calculated $T_{shock}$ at 4.3 mm.

The evolution of $T_{shock}$ with selected peak intensities (Fe I 382.8 nm and Fe II 238.3 nm) is shown in Figure 2(a) with $I_{actual}$ illustrated by dots and $I_{pre}$ by lines, where $q$=4. The consistency between $I_{actual}$ and $I_{pre}$ is presented in Figure 2(b) for Fe I 382.8 nm, in conjunction with 0.5707 RSD for $I_{actual}$ and 0.0178 RSD for the normalized $I_{actual}/I_{pre}$ (Figure 2(c)), which implies that PIGAs help decrease the RSD of the spectral intensity by 96.88%.

In order to validate the significance of reciprocal-logarithmic transformation, the other three with/without/semi-transformations are set up as control groups: $I$-$T_{shock}$(which original form has been commonly used in univariate or multivariate normalization strategies), ln$I$-$T_{shock}$, and $I$-$T_{shock}^{-1}$. In each group, all iron lines are calculated maintaining $q$=4 to get the predicted $I_{pre}$ and further normalize the actual fluctuating $I_{actual}$ by $I_{pre}$. We evaluate these four transformations in terms of $R^2$ and RE(%) between $I_{actual}$ and $I_{pre}$, and RSD of $I_{actual}/I_{pre}$ (Figure 2(d), (e) and (f)).

Box plots are created to present statistical properties of the above metrics where each group includes 108 Fe I lines and 33 Fe II lines. Each point represents one corrected result from one spectral line (center white line at the median, upper boundary at the 75th percentile, and lower boundary at the 25th percentile) with whiskers at ±1.5×interquartile range (IQR). ln$I$-$T_{shock}^{-1}$ performs truly the best with 0.9999 $R^2$, 1.074% RE(%) and 0.0134 RSD.

Another distinguishing feature of PIGAs has been found to be its suitability for spectral lines undergoing self-absorption, so let's demonstrate the possible reasons behind. It's worth mentioning that at the beginning of our derivation, $\tau$ is assumed to be zero. While for optically-thick plasma($\tau\neq0$) having the risk of reabsorbing photons, $I$ can be extended as

$$I(\nu)=B(\nu)\left(1-e^{-\tau(\nu)}\right)$$
$$=B\left(\tau-\frac{1}{2}\tau^2+\frac{1}{6}\tau^3-\frac{1}{24}\tau^4+\cdots\right)=-B\sum_{j=1}^{\infty}\frac{(-\tau)^j}{j!} \quad (7)$$

which means the spectral intensity of an optically-thick plasma can be expanded into a series of polynomials for the intensity $B\tau$ of the optically-thin one. It suggests that the proposed polynomial extension in PIGAs can somewhat compensate the simplification brought by the optically-thin assumption when we deal with the optically-thick plasma. The corrected RSD in Figure 2(f) reconfirms this conclusion, where self-absorption lines marked in brown, with 0.0168(median) still lower the uncorrected 0.5121(median) (grey dashed line).

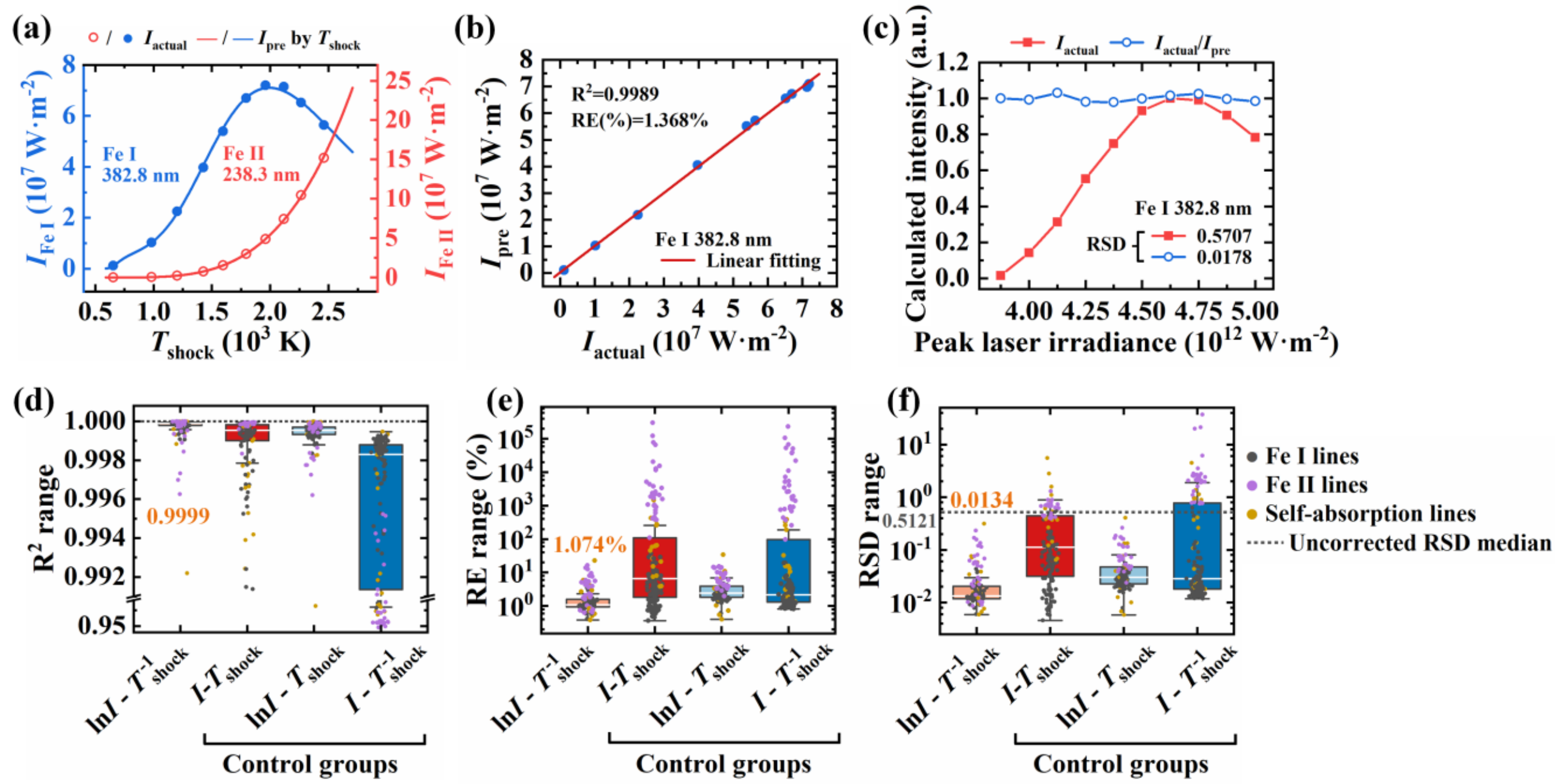

**FIG. 2** Correction results of PIGAs for calculated spectral lines through $T_{shock}$ at 4.3 mm. Evolution of (a)line intensities with $T_{shock}$ under all laser irradiances (dots from calculation, curves from PIGAs), (b)predicted $I_{pre}$ from PIGAs with the actual $I_{actual}$, (c)line intensities varying with irradiances before($I_{actual}$) and after($I_{actual}/I_{pre}$) correction. Box plots of (d)$R^2$, (e)RE(%) and (f)RSD of PIGAs and three other transformations for 108 Fe I lines and 33 Fe II lines.

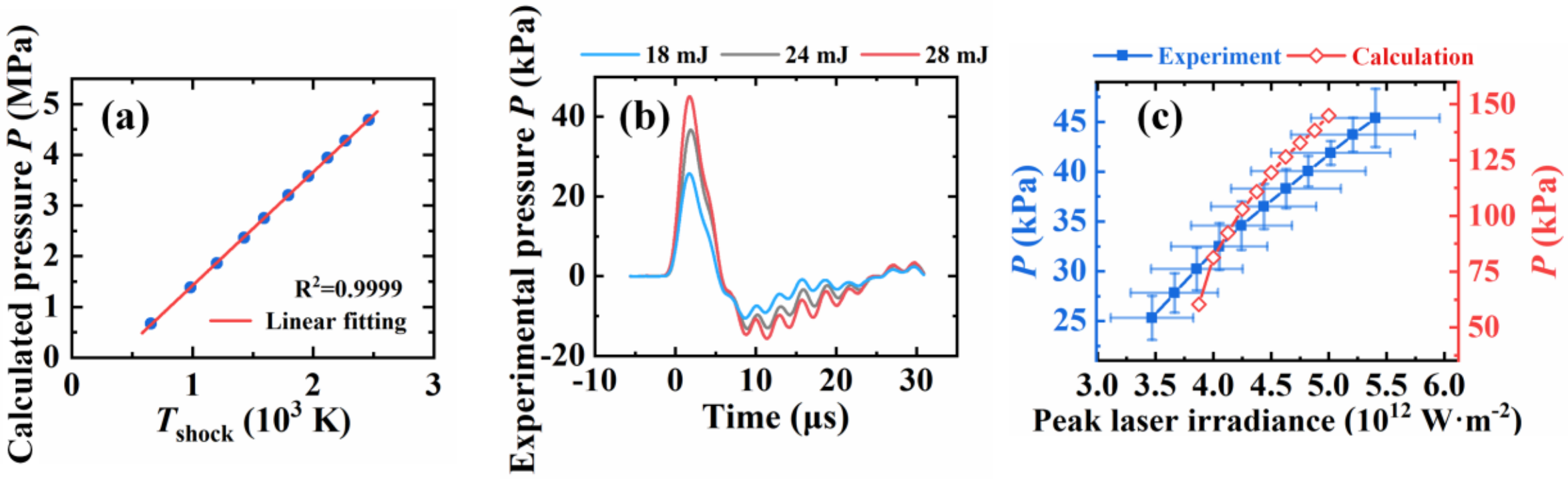

**FIG. 3** Evolution of the shockwave pressure at 4.3 mm. (a)The linear consistency between calculated $P$ and $T_{shock}$. (b)Temporal Evolution of the experimental shockwave pressure (propagation delay automatically returned to zero since being triggered by the shockwave itself). (c)Measured and calculated $P$ in different laser irradiances.

The experimental results are subsequently taken to validate PIGAs with the measured $P$ at 4.3 mm and the back-calculated $T_{shock}$.

The linear consistency between $P$ and $T_{shock}$ ($R^2$ reaching 0.9999) confirmed in ten calculated irradiances (blue dots in Figure 3(a)) guarantees the back-calculation. The measured $P$ is selected as the maximum value of the experimental shockwave profile (Figure 3(b)), which shows the same monotonic trend with laser irradiances as the calculated $P$ (calculated results filtered to match the 500 kHz resonance frequency limit of the pressure sensor in Figure 3(c)).

Figure 4(a) compares correction results of PIGAs ($q$=4) for Fe I 382.8 nm. Their RSDs goes from 0.2829 (uncorrected), 0.0234 (corrected by $P$) to 0.0170 (corrected by $T_{shock}$), which are reduced by 92% and 94%, respectively.

Subsequently we illustrate the specific fluctuating range for 108 Fe I lines (Figure 4(b)) and 33 Fe II lines (Figure 4(c)) before and after correction. Although $P$ and $T_{shock}$ are linearly correlated, the inconsistency in the occurrence of “peaks and valleys” in the corrected spectral intensities implies a nonlinear relationship between their corrected effects.

Specific RSDs results are shown in Figure 5. The median RSD of 108 Fe I lines transforms from 0.4663 (uncorrected), 0.0720 (corrected by $P$) to 0.0544 (corrected by $T_{shock}$), effectively reduced by 85% and 88% (Figure 5 (a)). While for 33 Fe II lines (Figure 5 (b)), it is reduced by 77% (corrected by $P$) and 86% (corrected by $T_{shock}$). Seventeen self-absorption lines (Fe I: Fe II=15:2) are also corrected effectively, with RSDs reduced by 78% (corrected by $P$) and 89% (corrected by $T_{shock}$) (RSD median: 0.3815→0.0825→0.0437), which are marked with gray reference lines

In conclusion, we present a normalization framework using shockwave characteristics ($P$ and $T_{shock}$) called PIGAs for LIBS spectral normalization. it features physical constraints on reciprocal-logarithmic transformation, genetic algorithms with multiple objectives optimizing unknown parameters in physical correlations, independence of fluctuation-inducing sources, and suitability to wide ranges of laser irradiance and spectral

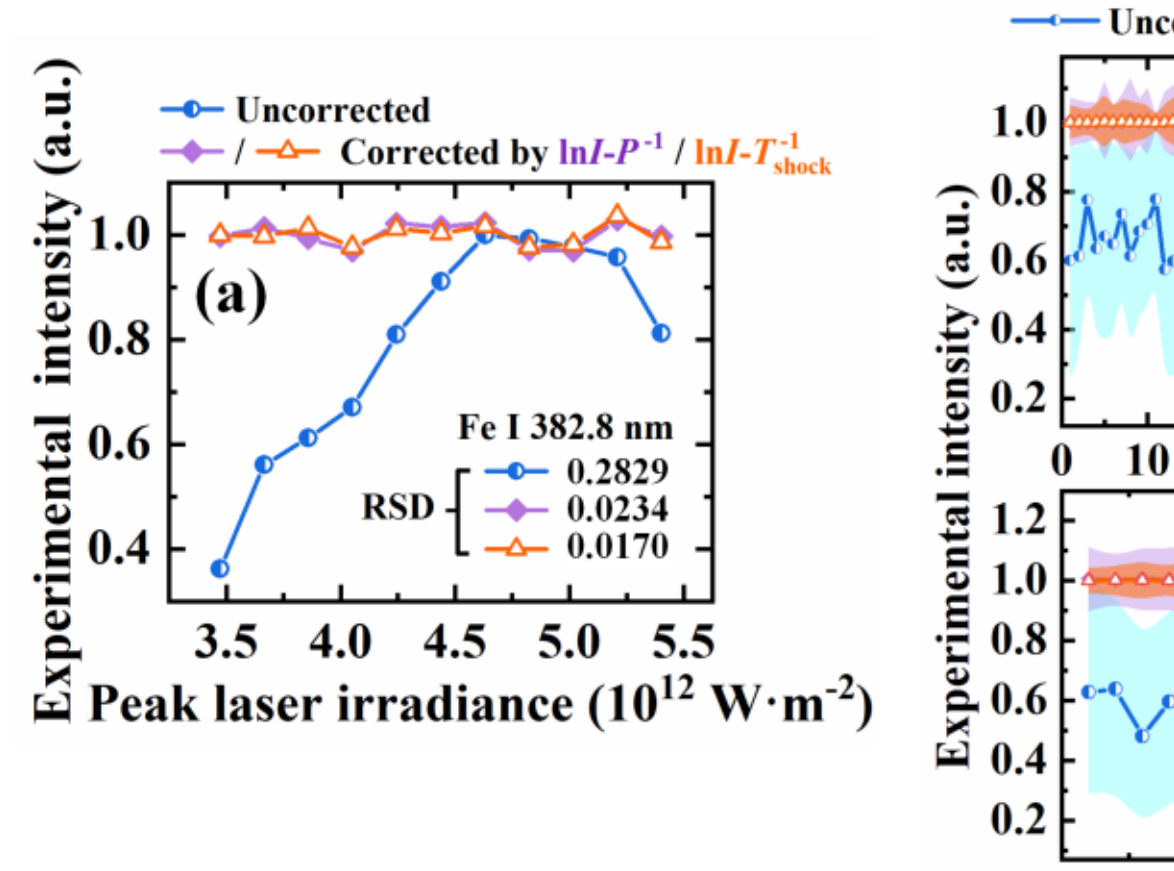

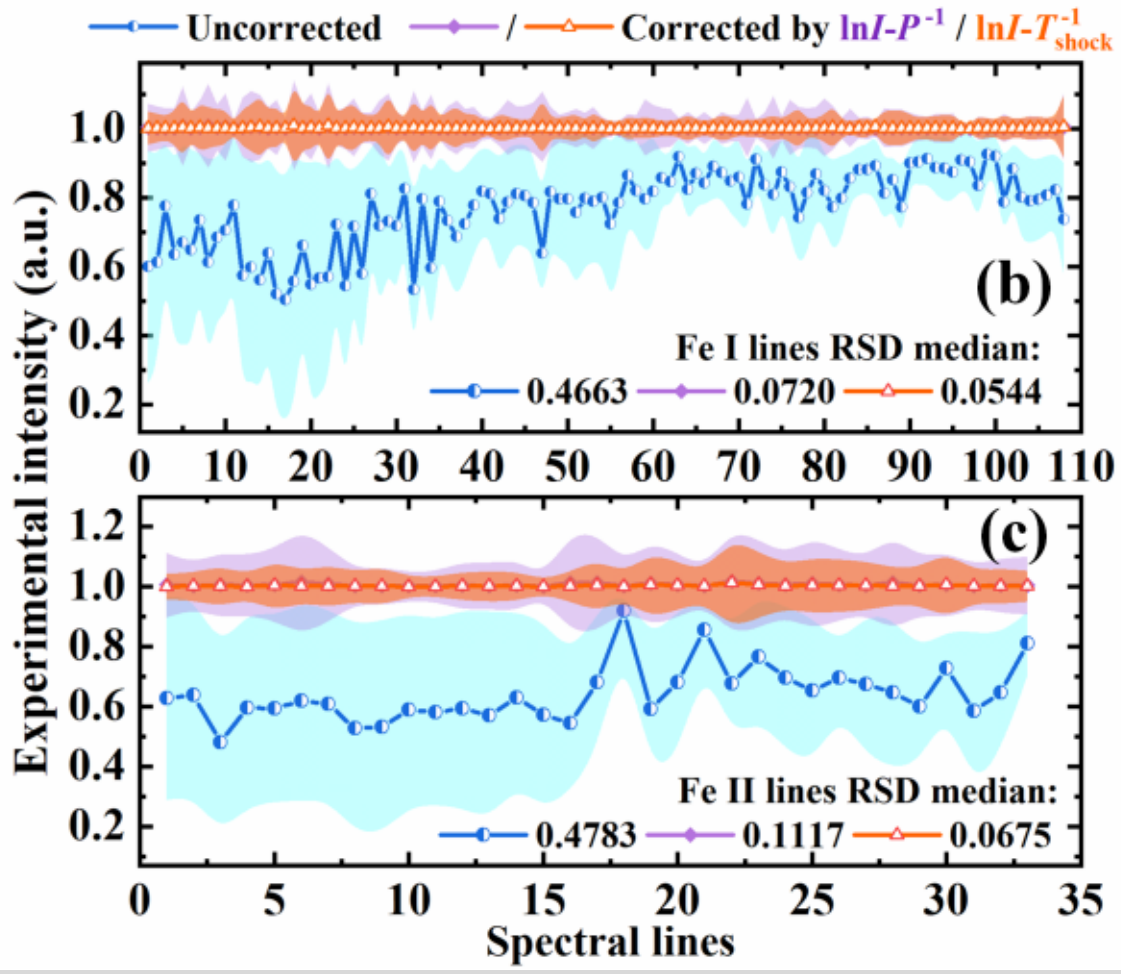

**FIG. 4** Experimental spectral normalization using PIGAs for (a)Fe I 382.8 nm, (b)108 Fe I lines and (c)33 Fe II lines.

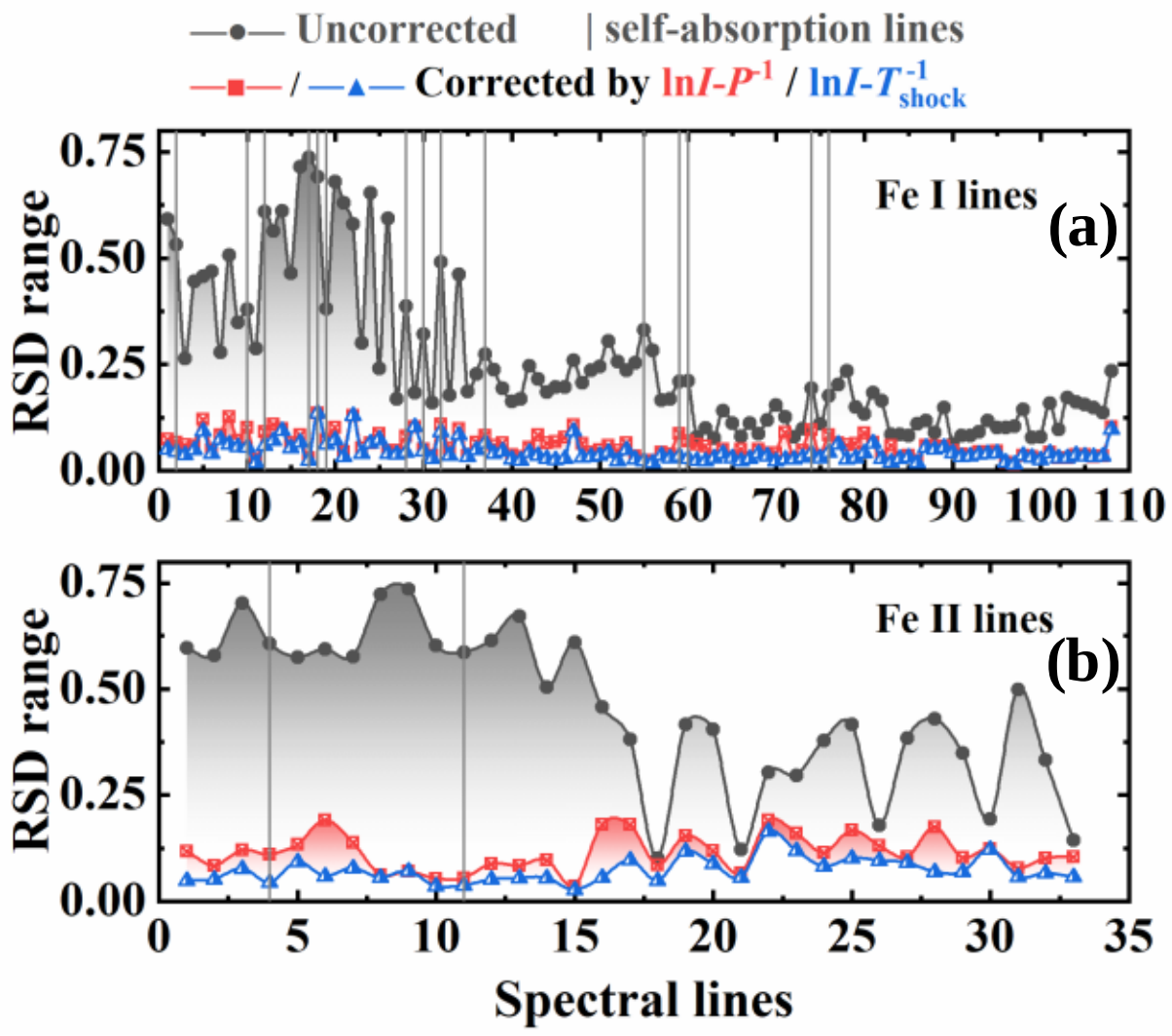

**FIG. 5** Specific RSDs of (a)108 Fe I lines and (b)33 Fe II lines before and after correction, with self-absorption lines marked by grey vertical lines.

lines undergoing self-absorption.

Ten laser irradiances covering laser-induced breakdown to significant plasma shielding are treated as one of the spectral fluctuation-inducing sources to demonstrate the correction effects of PIGAs with 108 Fe I and 33 Fe II lines. Calculated data from LAP-RTE codes are used as the benchmark to explain the physical plausibility of the reciprocal-logarithmic transformation and the applicability for self-absorbing spectral lines comparing with three other transformations. The median experimental RSD shows significant reductions of 85% (corrected by $P$) and 88% (corrected by $T_{shock}$) for 108 Fe I lines, 77% (corrected by $P$) and 86% (corrected by $T_{shock}$) for 33 Fe II lines. A certain corrective effect still works on 17 spectral lines undergoing self-absorption.

It should be noted that PIGAs is based on stoichiometric ablation, achieving a correlation and correction for the plasma temperature part in the spectral intensity. As for non-stoichiometric ablation or significant physical matrix effect, the particle number density $N$ part will no longer be proportional to the concentration $c_A$ within the sample, which may also fluctuate the following plasma parameters and further impair the quantitative analysis of the elemental calibration curve. These findings hope to provide some inspiring thinking also for other low-temperature plasmas with similar processes.

## SUPPLEMENTARY MATERIAL

See the supplementary material for the developed end-to-end LAP-RTE model, detailed derivation of the correlation between $T_e$ and $T_{shock}$, box plots of three metrics for experimental spectral normalization, correction for calibration curves for trace elements in iron-based alloys using PIGAs and the other normalization methods, and the set of lines considered in the calculation (Table S9).

## ACKNOWLEDGMENTS

This work was supported by the National Key Research and Development Program of China (Grant No. 2021YFB3700018) and Shaanxi Provincial Science and Technology Plan Project (2022TD-59). The authors acknowledge Dr. I.B. Gornushkin from BAM Federal Institute for Materials Research and Testing for many inspiring discussions about the plasma model.

## AUTHOR DECLARATIONS

### Conflict of Interest

The authors have no conflicts to disclose.

## DATA AVAILABILITY

The data that support the findings of this study are available from the corresponding author upon reasonable request.